\documentstyle[sprocl,axodraw]{article}

\def\be{\begin{equation}}
\def\ee{\end{equation}}
\def\bea{\begin{eqnarray}}
\def\eea{\end{eqnarray}}

\begin{document}


\title{SPONTANEOUS CP VIOLATION: ALTERNATIVE TO
THE STANDARD MODEL}

\author{Paul H. Frampton}

\address{
Institute of Field Physics, \\
Department of Physics and Astronomy,\\
University of North Carolina, CB \#3255,
Chapel Hill. NC 27599-3255.\\
E-mail: frampton@physics unc,edu}


\maketitle\abstracts{We propose a model of soft CP violation that evades the strong CP problem
  and can describe observed CP violation in the neutral kaon sector, both
  direct and indirect.  Our model requires two ``duark'' mesons carrying
  quark number two that have complex (CP-violating) bare masses and are
  coupled to quark pairs.  Aside from the existence of these potentially
  observable new particles with masses of several hundred GeV, we predict a
  flat unitarity triangle ({\it i.e.,} no observable direct CP
  violation in the $B$-meson sector) and a possibly anomalous branching ratio
  for the decay mode $K^+\rightarrow \pi^++\bar{\nu}\,\nu$.  }

\bigskip
\bigskip

\section{{\bf Introduction}}

\bigskip

I am happy to be invited at this Inaugural Conference of the Michigan Center for
Theoretical Physics, and to wish for the Center a glorious and distinguished future.

The standard model of particle physics involves three fermion families and
one Higgs doublet. Within this model, CP violation can manifest itself in
just two ways: through the complex phase $\delta$ in the Kobayashi-Maskawa
matrix\cite{KM}  and through the coefficient $\bar \theta$ of the Chern-Simons
term \cite{XA}. The complex Yukawa couplings of the Higgs boson can
contribute to of both these parameters, so that they would  be
expected to be comparable in magnitude.  However,
the standard model requires $\vert\delta\vert\sim 1$
 to describe CP violation for neutral kaons, yet it also requires
$\vert\bar \theta\vert \le 1.5\times 10^{-10}$ lest observable
 nuclear electric dipole moments be generated\cite{RG}. This
dramatic departure from naturality is the gist of the strong CP
problem, whose solution is a primary goal of this paper which is based
on work with Glashow and Yoshikawa\cite{FGY}.

There are other reasons to consider alternatives to the standard description
of CP violation: Recall that the CP-violation implicit in the standard model
does not seem to be sufficient to implement the prescient idea of Sakharov
through which the baryonic asymmetry of the universe may be
generated\cite{AS}.  Furthermore it would seem useful to have other models in
hand in the event that experiments now being carried out do not confirm the
standard-model prediction that $\sin{2\beta}= 0.7\pm 0.2$, where $\beta$ is
one of the vertex angles of the unitarity triangle. In this connection, we
note that unofficial averages of various experimental results, as well as
indirect theoretical arguments based on other available data\cite{XB},
{\it disfavor but do not yet exclude} our prediction that $\sin{2\beta}\simeq 0$.

Several simple fixes to the strong CP problem have been suggested.  The
simplest of these, a massless up quark or an invisible axion, are all but
excluded by observation and theoretical analysis. In more elaborate models,
CP is assumed to be violated spontaneously\cite{barr} (which usually leads to
unacceptable domain walls), or softly \cite{PF,BC,GG}. In the
latter models, various new heavy fermions and heavy bosons are introduced
with CP conservation imposed on all dimension-4 terms in the Lagrangian, but
not on the lower-dimension bare mass terms of the new particles. Most of
these models have been excluded by experiment: they are superweak mimics that
cannot reproduce the observed value of $(\epsilon'/\epsilon)_K$.

\bigskip

\section{{\bf The Duark Model}}

\bigskip

Our model of soft CP violation is simpler than its predecessors in that it
requires new bosons, but no new fermions. Let us begin with the Lagrangian
for a variant of the standard model where: (a) CP invariance is imposed on
all dimension-four terms; and (b) two Higgs multiplets are introduced.  The
latter hypothesis requires clarification.  A  discrete symmetry $\cal
D$ must be imposed on the Lagrangian to ensure that one Higgs multiplet
$H^u$ gives rise to up-quark masses, while the other $H^d$ gives rise to
down-quark masses.  A suitable  choice for $\cal D$ is an operation under which
all right-handed quarks and $H^u$ are odd, with all other fields even.  Note
that $\cal D$ invariance forbids Higgs mass terms proportional to
$H^{u\dagger}H^d$ which (if complex) would directly  contribute to
${\rm Arg\; Det}\; M$.  At this point
in the  explication of our model the Lagrangian is entirely CP
conserving, with $\bar\theta=0$ and a real orthogonal KM matrix\cite{FG}.

The essential extension  of the above-described Lagrangian  consists of
two
spinless bosons $\phi^{(a)} \ \ (a=1,2)$ that carry quark number two (or
baryon number ${2\over 3}$) and couple to quark pairs.  We assume that
each of the $\phi^{(a)}$
is a  color anti-triplet and weak $SU(2)$  singlet with electric charge
$Q={1\over 3}$. These particles are hereafter referred to as ``duarks.''
To specify the duark couplings, we denote the left-handed quark doublets
by $\Psi_L$ where:

\begin{equation}
\Psi_L=\pmatrix{u_L\cr Vd_L}
\end{equation}

\noindent
with color, flavor, and Dirac indices suppressed. The $u_L$ and $d_L$ are
left-handed quarks in the basis in which the tree-level mass matrices are
diagonal, and $V$ is the tree-level KM matrix, which is real and orthogonal
by hypothesis.  The duark couplings to quarks may be written:

\begin{equation}
f\,\phi^{(a)}_i\, \epsilon_{ijk}\, \epsilon_{bc}\,o^{(a)}_{nm}
\,\left(\tilde \Psi^{jbn}_L\,(i\gamma_0\gamma_2)\,\Psi_L^{kcm}
\right) + \rm h.c.
\end{equation}

\noindent where  the tilde
transposes the Dirac indices and the real antisymmetric matrix
$i\gamma_0\gamma_2$ produces a Lorentz scalar. Indices   $a=1,2$ label the
duarks, indices  $i,j,k$ label colors, indices  $b,c$ label weak isospin,
and indices $n,m$ label the flavors of the quark doublets.  The subscripted
$\epsilon$'s are the usual invariant antisymmetric matrices.  The constant
$f$ sets the overall scale of the couplings.  The $o_{n,m}^{(a)}$ are
$3\times 3$ real symmetric matrices in flavor space. All their entries are
assumed to be of order unity in lieu of specific theoretical insight.  In
brief, $\phi^{(a)}$ couples to the quark pair ($u_{Ln},\,d_{Lm}$) with
coupling strength $fo^{a}_{nl}\,V_{lm}$.  These duark-quark couplings are
necessarily flavor symmetric in the weak-isospin basis, but they depart from
symmetry in the mass-eigenstate basis we use.

The need  for two Higgs bosons now becomes apparent.
We have required  duarks to couple
  to pairs of left-handed quarks,
but not to pairs of right-handed quarks. This is a proper and renormalizable
  condition  if and only if we take   the $\phi^{(a)}$
to be even under the discrete symmetry $\cal D$.
(The other choice of $\cal D$-parity for duarks leads to a model with a
strong CP problem.)
We note in passing that $\cal D$ is broken spontaneously along with
$SU(2)\times U(1)$, so that finite  duark couplings to right-handed quarks
 arise at one loop. These are suppressed by a product of Higgs coupling
constants and by a canonical factor of $(4\pi)^{-2}$ and are small enough to
have no effect on our subsequent arguments.

Aside from their quartic self couplings (which play no role here), the duarks
have bare masses  (specified  by ${\cal M}^2$) and quartic
couplings to the Higgs bosons:

\begin{equation}
\phi^{(a)\,\dagger}\phi^{(b)}\,{\cal  M}^2_{(ab)} +
f^2\,\phi^{(a)\,\dagger}\phi^{(b)} \left\{
\alpha^u_{(ab)}\,H^{u\,\dagger} H^u + \alpha^d_{(ab)}\,H^{d\,\dagger}
H^d\right\}
\end{equation}

\noindent
The latter interaction is taken to have a coupling strength of order $f^2$, a
hypothesis both  suggested by and compatible with the assumed strength of
the duark Yukawa couplings to quarks.  The matrices $\alpha^{u,d}$ are real
and symmetric (CP conserving) with entries, once again, assumed to be of
order unity.  The bare masses of the duarks are described by the complex
Hermitean matrix ${\cal M}$. Indeed, the sole source of CP violation in
our model lies in the bare masses of the duarks.

The duark mass eigenstates  $\Phi^{(a)}$ are the eigenvectors of their
complete  mass matrix ${\cal M}^{\prime 2}$, which is the sum of
${\cal M}^2$ and
small additional terms obtained by replacing the Higgs bosons by their vev's
in Eq.~3. We denote these eigenstates by $\Phi= u^\dagger \phi$, where $u$ is
a $2\times 2$ unitary unimodular matrix.  Both mass eigenvalues $M^{(a)}$ and
their difference are assumed to be of comparable magnitude and are  denoted
by $M$.  This crude approximation (and the other simplifications we have
made) compel us to remind the reader that the estimates we shall offer for
duark masses and couplings are merely  order of magnitude estimates.

We rewrite the duark couplings given by Eq.~2 in terms of these mass
eigenstates. That is, we replace $\phi^{(a)}$ by $\Phi^{(a)} $ and the real
symmetric matrices $o^{(a)}$ by the complex symmetric matrices $O^{(a)}\equiv
o^{(b)}\,u_{ba}$.  In this manner, the CP violation is transferred from the
mass terms of the duarks to their couplings with quarks.  The complex (CP
violating) phases of these couplings are assumed to be of order unity.

\bigskip

\section{{\bf CP Violation in The Neutral Kaon Sector }}

\bigskip

As in any model of soft CP violation,
the box diagram shown in Fig.~1, with two incoming strange quarks and two
exiting down quarks, generates an effective 4-fermion coupling
of the form $(\bar d_L\gamma^\mu s_L)\,(\bar d_L\gamma_\mu s_L)$
which must be
wholly responsible for indirect CP violation in the neutral kaon sector.
\begin{center}
\begin{picture}(300,200)(0,0)
\SetWidth{1.2}
{\LARGE
\ArrowLine(100,25)(50,25)
\ArrowLine(100,25)(200,25)
\ArrowLine(250,25)(200,25)
\ArrowLine(50,125)(100,125)
\ArrowLine(200,125)(100,125)
\ArrowLine(200,125)(250,125)
\DashArrowLine(100,125)(100,25){5}
\DashArrowLine(200,25)(200,125){5}
\Text(50,130)[b]{$ s_L $ }
\Text(150,130)[b]{$ u,c,t $}
\Text(250,130)[b]{$ d_L $}
\Text(250,22)[t]{$ s_L $ }
\Text(150,22)[t]{$ u,c,t $}
\Text(50,22)[t]{$ d_L $}
\Text(95,75)[r]{$ \Phi $}
\Text(205,75)[l]{$\Phi $}}
\end{picture}\\
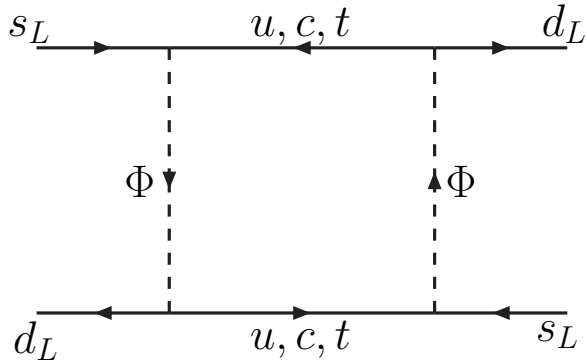
{\bf Fig. 1. Box diagram}
\end{center} 

\bigskip
\bigskip
\bigskip

\noindent Setting its coefficient equal to the experimentally determined value of
$\epsilon\, \Delta m_K$, one   obtains \cite{GG} the constraint:

\begin{equation}
{\alpha_f\over M} \approx 2\times 10^{-8}\,\rm GeV^{-1}
\end{equation}

\noindent
with $\alpha_f=f^2/4\pi$ and $M$  an estimate of the duark mass scale.
\smallskip

We turn to the question direct CP violation in the kaon sector,
such as discussed in \cite{AFKL,FG,GG}. Here our
model differs radically from its predecessor: the exchange of a duark between
two quark pairs,
 as shown in Fig.~2, generates
small and non-conventional  four-fermion couplings that
 contribute to non-leptonic decays.
\begin{center}
\begin{picture}(300,200)(0,0)
{\LARGE
\SetWidth{1.2}
\ArrowLine(250,25)(50,25)
\ArrowLine(50,125)(120,125)
\ArrowLine(250,125)(120,125)
\ArrowLine(160,75)(250,95)
\ArrowLine(160,75)(250,55)
\DashArrowLine(120,125)(160,75){5}
\Text(50,130)[b]{$ s_L $}
\Text(250,130)[b]{$ u_L $}
\Text(250,99)[b]{$ d_L $}
\Text(250,50)[t]{$ u_L $}
\Text(50,40)[t]{$ d  $}
\Text(150,105)[b]{$ \Phi $} }
\end{picture}\\
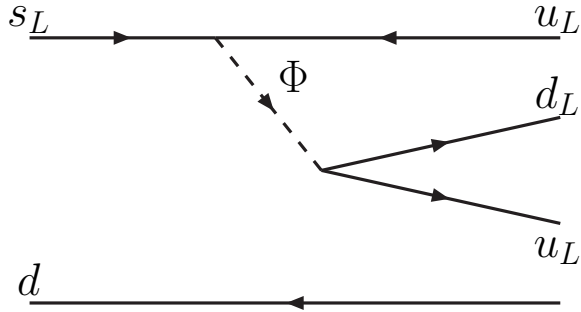
{\bf Fig. 2 Tree level contribution to direct CP violation}
\end{center}

\bigskip
\bigskip
\bigskip

These are of no significant observable consequence, except for the case of
neutral kaons where the  relevant term
 $(\bar d_L\gamma_\mu u_L)(\bar u_L\gamma^\mu s_L)$
is purely $\Delta I={1\over 2}$, has  magnitude $\sim f^2/M^2$ with a
large but unknown complex phase $\eta$. (It also has
 an unconventional color structure.) This term
  contributes comparably to the 2-pion decays of both $K_L$ and $K_S$, so
 that the overall  decay amplitudes
will satisfy:

\begin{equation}
A_2/A_0\Big\vert_S=\omega\,(1-\zeta\cos{\eta} -i\epsilon\zeta \sin\eta)
\end{equation}
\noindent and
\begin{equation}
A_2/A_0\Big\vert_L=\omega\,(1-\zeta\cos{\eta} - i(\zeta/\epsilon)\sin\eta)\,,
\end{equation}

\noindent where $\omega\equiv A_2/A_0\vert_S\simeq 1/22$,  $\zeta$ is
the ratio  of the strength of the duark exchange amplitude to that
arising from $W$ exchange, and $\eta$ is its unknown phase:

\begin{equation}
\zeta \approx {f^2\over M^2} \big(\sqrt{8}\,G_F\sin{\theta_c}\big)^{-1}
\end{equation}

\noindent with $\theta_c$ is the Cabibbo angle. Here
we ignore the difference in color structure of the two amplitudes.

From these results (and the known relation among pion phase shifts,
$\delta_2-
\delta_0\approx -\pi/4$)  we deduce:

\begin{equation}
1-6 {\epsilon'\over \epsilon}\equiv
\left\vert {\eta^{+-}\over \eta^{00}}\right\vert^2= 1\pm
{3\zeta\omega\over
   \epsilon}\,,
\end{equation}

\noindent
where we have arbitrarily chosen $\eta=\pm \pi/2$ for the unknown phase.
>From this result  we obtain

\begin{equation}\left\vert{\epsilon'\over \epsilon}\right\vert
\approx  \left\vert {\zeta\omega\over
 2  \epsilon}\right\vert\,.
\end{equation}

\noindent
If Eq.(8) is to yield the observed value
$\epsilon'/\epsilon\simeq 2\times
  10^{-3}$,  we must have   $\zeta\approx 1.8\times
10^{-4}$. Making use of this result and Eq.~6, we
  obtain the following order of magnitude
constraint:

\begin{equation}
   {\alpha_f\over M^2}\approx 10^{-10}\  \rm GeV^{-2}
\end{equation}

\noindent which together with Eq.~4,
the constraint  from indirect CP violation,
yields the estimates
\begin{equation} M\approx 200~{\rm GeV \quad\ and}\quad\  \alpha_f\approx 4\times10^{-6}\,.  
\label{ests} 
\end{equation} 
It must be emphasized that, because of the assumptions used, Eq.(\ref{ests}) 
gives only order of magnitude estimates: 
the mass M is predicted to be several hundred GeV.  
We have shown that a correct description of CP violation 
in the neutral kaon system can be obtained with duarks of 
soon-to-be-accessible masses.  However, our model requires 
that duarks are only weakly coupled to quark pairs.  

\bigskip 

\section{{\bf Direct CP Violation in $B$-Meson Decays}}

\bigskip

Our model starts off with a real KM matrix and a degenerate unitarity
triangle.  Radiative corrections will produce a finite imaginary part of the
KM matrix, with the leading contribution arising from the Feyman diagram
shown in Fig. 3. 
\begin{center}
\begin{picture}(300,200)(0,0)
\SetWidth{1.2}
{\LARGE
\ArrowLine(50,50)(90,50)
\ArrowLine(210,50)(90,50)
\ArrowLine(210,50)(250,50)
\DashArrowArcn(150,50)(60,180,0){5}
\Text(50,46)[rt]{$\Psi_L^n $}
\Text(150,46)[t]{$\tilde{\Psi}_L^c $}
\Text(250,46)[lt]{$\Psi_L^m $}
\Text(150,118)[b]{$\Phi $ } }
\end{picture}\\
{\bf Fig. 3 Self-mass.}
\end{center}

\bigskip
\bigskip
\bigskip

The situation here is similar to that in earlier models of
soft CP violation, where the area of the unitarity triangle divided by its
standard-model value is typically $\sim\!\alpha_f/4\pi$, where $\alpha_f$
characterizes the couplings of hypothetical new particles to quarks.  In our
case, this coupling constant is tiny and the unitarity triangle remains
experimentally indistinguishable from a straight line.  The exchange of a
duark between quark pairs, as shown in Fig.~2, does yield  a non-standard
contribution to  $B$-decay, but one which is  three orders of
magnitude smaller than that due to $W$ exchange.  It does not lead to
readily detected effects.  We conclude that our model demands a flat
unitarity triangle and predicts that there is no observable direct CP
violation in the $B$-meson sector.

\bigskip

\section{{\bf The Strong CP Problem}}

\bigskip

In our model, as in all models of soft CP violation,  the QCD $\theta$
parameter---corresponding to a dimension-4 CP violating operator---must
 vanish. All CP-violating contributions to $\bar\theta$ are
finite and calculable radiative corrections. We need not consider self-energy
diagrams such as that in Fig.~3  because they are associated with Hermitean
counter-terms in the Lagrangian and cannot contribute to $\bar\theta$ to any
order in $\alpha_f$.
Rather, and as further explicated  in \cite{GG}, we must examine radiative
corrections to the Higgs couplings, which is to say, corrections to quark
masses ($\Delta M_U$ and $\Delta M_D$) whose contribution to the phase of the
determinant of the quark masses is:

\begin{equation}
\Delta\, \bar\theta \approx \rm Im\; Tr\left(\Delta M_D\,M_D^{-1}\right)+
Im\; Tr \left(\Delta M_U\,M_U^{-1}\right)
\end{equation}

 \noindent  Earlier models of soft CP violation involve the existence of new
 heavy particles (both mesons and fermions) which are very large compared to
 the electroweak scale.  In contrast, our new particles (duark mesons) have
 relatively small masses.  Thus, quark masses appearing in the denominators
 of Feynman integrals contributing to $\Delta M_U$ and $\Delta M_D$ cannot be
 ignored. However, this complication is alleviated by the tiny value we have
 deduced for $\alpha_f$. It is sufficiently small that we need examine only
 those quark mass corrections involving exactly one duark loop. The leading
 contributions to $\bar\theta$ arise from diagrams with one duark loop and
 one Higgs loop, for which there are two possibilities.

 One of the leading contributions to $\bar\theta$ in our model arises from
 the two-loop diagram shown in Fig.~4a.  
\begin{center}
\begin{picture}(450,250)(0,0)
\SetWidth{1.2}
{\LARGE
\ArrowLine(50,100)(100,100)
\ArrowLine(160,100)(100,100)
\ArrowLine(200,100)(160,100)
\ArrowLine(240,100)(200,100)
\ArrowLine(240,100)(300,100)
\ArrowLine(300,100)(350,100)
\DashArrowArcn(170,100)(70,180,0){5}
\DashArrowArc(230,100)(70,180,360){5}
\Text(50,96)[rt]{$c_L$}
\Text(130,96)[t]{$b_L$}
\Text(180,96)[t]{$t_R$}
\Text(220,96)[t]{$t_L$}
\Text(280,96)[t]{$b_L$}
\Text(350,96)[lt]{$c_R$}
\Text(200,100)[]{$\times $}
\Text(203,110)[b]{$ m_t $ }
\Text(165,162)[t]{$\Phi $ }
\Text(240,39)[b]{$H^+$ }}
\end{picture}\\
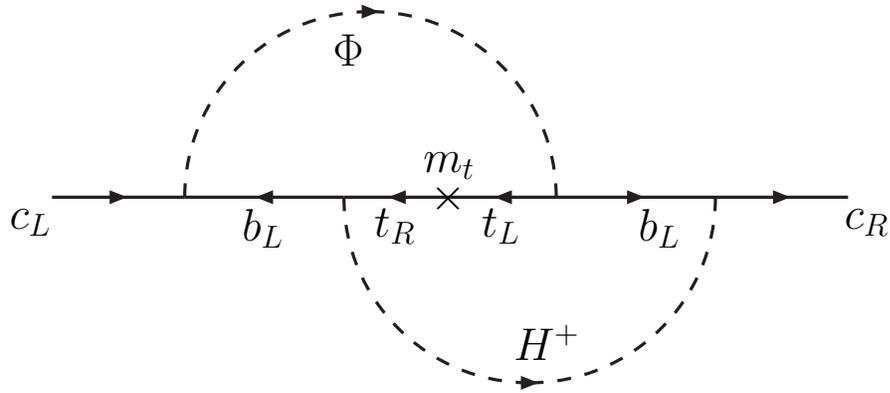
{\bf Fig. 4a First 2-loop contribution to $\bar{\theta}$ discussed in the text.}
\end{center}

\begin{center}
\begin{picture}(450,250)(0,0)
\SetWidth{1.2}
{\LARGE
\ArrowLine(50,50)(110,50)
\ArrowLine(200,50)(110,50)
\ArrowLine(200,50)(290,50)
\ArrowLine(290,50)(350,50)
\DashArrowArcn(200,50)(90,180,90){5}
\DashCArc(200,50)(90,0,90){5}
\DashLine(200,140)(200,170){5}
\DashArrowLine(200,140)(200,50){5}
\Vertex(200,140){3}
\Text(55,45)[t]{$d_L $}
\Text(150,45)[t]{$u_L $}
\Text(250,45)[t]{$d_L $}
\Text(350,45)[t]{$ d_R $ }
\Text(120,100)[rb]{$\Phi $ }
\Text(195,100)[r]{$\Phi $ }
\Text(285,100)[lb]{$H $ }
\Text(200,170)[b]{$<H>$}}
\end{picture}\\
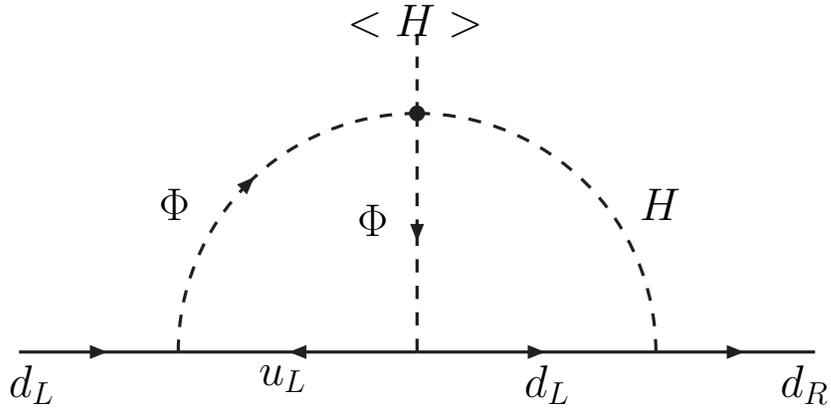
{\bf Fig. 4b Second 2-loop contribution to $\bar{\theta}$ as discussed in the text.}
\end{center}

\bigskip
\bigskip
\bigskip

The imaginary part of this amplitude
 tends to zero as the KM matrix $V$ approaches the unit matrix.  It also
 tends to zero if all quark masses (except the one with the mass insertion)
 are neglected.  It follows that these contributions to $\bar\theta$ are
 highly suppressed:

\begin{equation}
\Delta\,\bar\theta \simeq {\alpha_f\,\lambda^2\over (4\pi)^3}\,\left\{
{m_t^2\,m_b^2\over M^2\,\langle H^u\rangle^2}\quad\ {\rm or} \quad\
{m_b^2\over \langle H^d\rangle^2}\right\}
\end{equation}

\noindent where $\lambda$ is the Wolfenstein parameter
whose value is approximately $\sin{\theta_c}$. These
radiative corrections  yield
 $\Delta \bar\theta \sim 10^{-12}$.
 The other  two-loop  contribution to $\bar\theta$, shown in
Fig.~4b, involves the quartic Higgs-duark coupling of Eq.~3. This
radiative correction contributes  $
\Delta\bar\theta\sim (\alpha_f^2/4\pi)^2\simeq 10^{-13}$.
It follows that our model is easily compatible with present constraints on the
strong CP-violating parameter $\bar\theta$.

\bigskip

\section{{\bf Rare Semi-Leptonic Kaon Decays}}

\bigskip

Here we consider the rare decay modes $K^+\rightarrow \pi^+ \bar{\nu}\,\nu$
and $K_L\rightarrow \pi^0 \bar{\nu}\,\nu$. The first of these has a predicted
branching ratio of $(7.9\pm 3.1)\times 10^{-11}$ ~~~\cite{buras}.  The observed
branching ratio of $(15^{+34}_{-12}) \times 10^{-11}$ ~~~\cite{adler} agrees with
the prediction, but is based on the observation of a single event at the
Brookhaven E787 experiment.

In our model, $Z^0$ penguin diagrams involving a duark loop contribute to
both of these decay modes. The dominant contribution to this diagram (as for
the standard-model penguin) involves an intermediate top quark.
Thus for
    the decay $K^+\rightarrow \pi^+\bar{\nu}\,\nu$,
the ratio of this novel
 amplitude to the conventional amplitude is  given naively by

\begin{equation}
\zeta' =   {f^2\over M^2}\,\left(\sqrt{8}G_F\,\lambda^5\right)^{-1}\simeq
\ 0.1
\end{equation}

\noindent
Of course, this   is no better than an order-of-magnitude
 estimate. In fact, $\zeta'$ may be negligible, or it may be of order unity.
Thus  future measurements of rare semileptonic kaon decays
 may reveal a significant departure from  standard model predictions.
 A more careful  calculation of the duark
contribution (such as \cite{buras} in the case of the standard-model result)
is certainly premature at present, but
will  be called for if more precise data becomes available, and if our model
 of soft CP violation survives further experimental scrutiny.

\bigskip

\section{{\bf Conclusion}}

\bigskip

We sketched a model wherein CP is a softly-broken symmetry of the Lagrangian.
Our  model yields $\vert\bar\theta\vert<10^{-12}$ and therefore
does not suffer from a  strong CP problem.
It is based on a two-Higgs variant of the
standard model to which are adjoined two spinless duark mesons
that carry baryon number $2\over 3$ and have  CP
violating bare masses.\footnote{Fanciers of supersymmetry may wish
  to identify anti-duarks with squarks that enjoy $R$-odd couplings
 that
  violate baryon number by one.}
  Indirect CP violation in
kaon decay ($\epsilon$ related) proceeds through a duark box diagram.  Unlike
other models of soft CP violation, ours includes direct CP violation via duark
 that readily accommodates the observed value of $\epsilon'$.
It would be interesting to study the incorporation of our model in
a larger theoretical scheme such as grand unification or higher dimensions.

Furthermore.  our model makes two immediate and decisive predictions: There
should be no observable direct CP violation in the $B$-meson sector. That is,
we predict $\sin{2\beta}=0$ to the precision of any currently feasible
experiment.
 Secondly, we predict the existence of duarks with masses that
are soon to be experimentally accessible.  These particles should be
copiously pair-produced at the LHC with picobarn cross sections, and they
should decay into assorted quark pairs (tb, td, bc, etc.) with widths well
below experimental resolution. Those
events involve a $t\,\bar t$ pair and two
additional  jets should provide a recognizable signal.
 In addition to these explicit predictions, we
find that penguin diagrams involving duarks may contribute significantly
to decays of kaons into pions plus $\nu\,\bar\nu$ pairs.

\bigskip
\bigskip
\bigskip

\bigskip

\section*{{\it Acknowledgements}}

The researches of PHF and TY
were  supported in part by the US Department of Energy
under Grant No. DE-FG02-97ER-41036.


\bigskip
\bigskip
\bigskip




\section*{References}
\begin{thebibliography}{99}
\bibitem{KM}
M. Kobayashi and T. Maskawa, Prog. Theor. Phys. {\bf 49,} 652 (1973).
\bibitem{XA}
G. 'tHooft, Phys. Rev. Lett. {\bf 37,} 8 (1976)\\
R. Jackiw and C. Rebbi, Phys. Rev. Lett. {\bf 37,} 172 (1976)\\
C.G. Callan, R.F. Dashen, and D.J. Gross, Phys.  Lett. {\bf B63,} 334
(1976).
\bibitem{RG}
M.V. Romalis, W.C. Griffith
and E.N. Fortson, {\tt hep-ex/0012001.}
\bibitem{FGY}
P.H. Frampton, S.L. Glashow and T. Yoshikawa, Phys. Rev. Lett. {\bf 87,} 011801 (2001).
\bibitem{AS}
A.D. Sakharov, JETP Lett. {\bf 5,} 24 (1967).
\bibitem{XB}
A. Abashian, {\it et al}. (BELLE Collaboration).
Phys. Rev. Lett. {\bf 86,} 2509 (2001).\\
B. Aubert, {\it et al}. (BABAR Collaboration).
Phys. Rev. Lett. {\bf 86,} 2515 (2001).
\bibitem{barr}
S.M. Barr, Phys. Lett. {\bf B448,} 41 (1999);\\
See also: S.M. Barr and A.Zee, Phys. Rev. Lett. {\bf 55,} 2253 (1985);
S.M. Barr, Phys. Rev. {\bf D34, } 1567 (1986);
S.M. Barr and E.M. Freire Phys. Rev. {\bf D41,} 2129 (1990).
\bibitem{PF}
P.F. Frampton and T.W. Kephart, Phys. Rev. Lett. {\bf 66,} 1666 (1991),\\
P.H. Frampton and M. Harada, Phys. Rev. {\bf D59,} 036004 (1999).
\bibitem{BC}
D. Bowser-Chao, D. Chang and W.Y. Keung, Phys. Rev. Lett. {\bf 81,} 2028
(1998) and Phys. Rev. {\bf D59,} 035004 (1999).
\bibitem{GG}
H. Georgi and S.L. Glashow, Phys. Lett. {\bf B451,} 372 (1999).
\bibitem{FG}
P.H. Frampton and S.L. Glashow, Phys. Rev. {\bf D55,} 1691 (1997).
\bibitem{AFKL}
A.W. Ackley, P.H. Frampton, B. Kayser and C.N. Leung,
Phys. Rev. {\bf D50,} 3560 (1994).
\bibitem{buras}
A.J. Buras. {\tt hep-ph/9905437.}
\bibitem{adler}
S. Adler {\it et al}~~ Phys. Rev. Lett. {\bf 84,} 3768 (2000)
\end{thebibliography}
\end{document}